\documentstyle[prc,aps,epsf,epsfig,amssymb,amsbsy,amstext,subeqnarray,amsfonts]{revtex}
\begin{document}

\title{Fluctuations of the quark densities in nuclei}

\author{G. Chanfray\protect\( ^{1}\protect  \), M. Ericson\protect\(
^{1,2}\protect  \)}

\address{{ \protect\( ^{1}\protect \)IPN Lyon, IN2P3-CNRS et UCB
 Lyon I, F69622 Villeurbanne Cedex} }

\address{{ \protect\( ^{2}\protect \)Theory division, CERN, CH12111 Geneva} }

\maketitle

\begin{abstract}
We study the static scalar susceptibility of the nuclear medium, {\it i.e.}, the
change of the quark condensate for a small modification of the quark mass.
In the linear sigma model it is linked to the in-medium  sigma propagator and its
magnitude increases due to the 
mixing with the softer  modes of the nucleon-hole excitations.
We show that the pseudoscalar susceptibility, which is large in the vacuum, 
owing to the smallness of the pion mass, follows the density evolution of 
the quark condensate and thus decreases.  At normal nuclear 
matter density  the two susceptibilities become much closer, 
a partial chiral symmetry restoration effect   as they become 
equal when the full restoration is achieved. 
\end{abstract}

 \section{Introduction}

The problem of the quark condensate and chiral symmetry restoration
in a dense medium has been extensively addressed. Little
attention
instead has been given to the question of the fluctuations and
correlations of the quark scalar density.
However for what concerns  the chiral phase transition this
quantity  is as relevant as the evolution of the quark condensate. 
In the case of a second order phase transition, large spontaneous 
fluctuations occur with an increase of the correlation length and of the
relaxation time. 

\smallskip\noindent 
Nuclear matter at normal density is well  below the critical density.
Nevertheless chiral symmetry is appreciably restored. The order
parameter decreases, as compared to its vacuum value, by $\simeq 35\%$. 
One may then wonder if there exist also large spontaneous  fluctuations
of the quark scalar density.
The present work addresses these questions. Our investigation is performed  
in an effective theory, the linear sigma model.  The two fields 
introduced in this model to insure chiral symmetry, the pion and the sigma,
have a great relevance in nuclear physics, in such a way that we can use,
as a source of information, the experience acquired in this field. We
investigate in
particular the correlations of the quark scalar density, showing how
their range increases with density. We also introduce the fluctuations of 
the pseudoscalar quark density. We  investigate how it evolves with 
density and its relation to the order parameter.


 \section{Scalar Susceptibility}

For an infinite system which possesses translational invariance the
quark density correlator  only depends on the relative space-time
separation $x$. It can be defined as the retarded Green's function  
$\langle -i\Theta(x^0)
\big[\overline{q}q(x),\overline{q}q(0)\big]\rangle$.
Previous investigations  \cite{CSSS95,DZCW00} 
have addressed the question of the in-medium four-quark condensate,
{\it i.e.}, the
quantity  $\langle \overline{q}q(x)\overline{q}q(x)\rangle$ which is  a
correlator taken at the same space-time point.

\medskip\noindent   
In the linear sigma model the symmetry breaking piece of the
Lagrangian is proportional to the sigma field~:
\begin{equation}
{\cal L}_{\chi SB}=\,c\,\sigma
\end{equation}
with $c= f_\pi\,m_\pi^2$. This quantity
 plays the role of the symmetry breaking Lagrangian of QCD~:
\begin{equation}
{\cal L}_{\chi SB}^{QCD}=- 2\,m_q\,\bar q\, q
\end{equation}
\noindent 
where $\bar q q=(\bar u u+\bar d d)/2$ , $m_q=(m_u+m_d)/2$ and we have
neglected isospin violation.
Making use of the Gell-mann-Oakes-Renner relation we obtain  the
following correspondence between the QCD and  effective theory
correlators:
\begin{equation}
{\langle\bar q q(x)\,\bar q q(0)\rangle\over \langle\bar q
q\rangle_{vac}^2}=
 {\langle \sigma(x)\,\sigma(0)\rangle\over f^2_\pi}
\end{equation}  
where  $\langle qq\rangle_{vac}$ is the vacuum value of the condensate. The
fluctuations of the quark density are thus carried by the sigma field, the
chiral partner of the pion. 
The in-medium  propagation of the sigma  in the energy domain near the 
two-pion threshold, has been 
the object of several investigations (see e.g. \cite{ARGCW95,HKS99,WAS01}).  Here
we will focus on aspects which have been ignored,
namely the low energy region, below the particle-hole excitation energies. 

\smallskip\noindent
The increase in the range of the correlator is reflected in the increase of
the static susceptibility which can become divergent at the 
critical density in case of a second order phase transition,  
a consequence of the appearance of a soft scalar mode \cite{HK85,RW93}.
In QCD the conjugate variables are the quark scalar density, which is the
order parameter, and the quark mass, which is the exciting field,
analogous to the magnetic field. 
As an amusing illustration of this analogy, in 
the Nambu-Jona-Lasinio (NJL) model, the constituant quark mass
differs from the current one by the effect of the interaction with
the condensate~:
\begin{equation}
M_q=m_q\,-\,2\,G_1\,\langle \bar q q\rangle
\end{equation}
This relation presents an analogy with that between  the magnetic field
inside the ferromagnet and the applied one in the Weiss
theory of magnetism~:
\begin{equation}
{\bf H}={\bf H}_0\,+\,\lambda\,{\bf M}
\end{equation}
These two quantities differ by the existence of an internal field, $\lambda\,{\bf M}$, 
proportional to the magnetization ${\bf M}$. In NJL, the action of the quark
condensate on the quark  mass thus plays the role of the
internal field of the Weiss theory.

\smallskip\noindent
The scalar susceptibility of QCD per unit volume represents the modification of 
the order parameter, which is the quark 
condensate,  to a small perturbation of the quark mass, the parameter responsible 
for the explicit symmetry breaking:
\begin{equation}
\chi_S={\partial \langle\bar q q\rangle\over\partial m_q}= 2\,
\int \,dt'\,d{\bf r}'\, G_R({\bf r}=0, t=0\, ,\,{\bf r}', t')
\end{equation}
where $G_R$ is the retarded quark scalar correlator~:
\begin{equation}
G_R({\bf r}, t, {\bf r}', t')=\Theta(t-t')\,
\langle-i \big[\bar q q({\bf r}, t)\, , \,\bar q q({\bf r}', t')\big]\rangle
\end{equation}
The  susceptibility represents space and time integrated correlators. 
In the effective theory, the quark density fluctuations 
are carried by the sigma field and the corresponding  scalar static
susceptibility is given by~:
\begin{eqnarray}
\chi_S &=& \,2\,{\langle \bar q q\rangle^2_{vac}\over f^2_\pi}
\, \int_0^\infty
\,d\omega\left({2\over\pi\omega}\right)\,Im D_{SS}({\bf q}=0,\omega)\nonumber\\
&=&\,2\,{\langle \bar q q\rangle^2_{vac}\over f^2_\pi}\, 
Re D_{SS}({\bf q}=0,\omega=0)
\end{eqnarray}
where $D_{SS}({\bf q},\omega)$ is the Fourier transform of the scalar
correlator~:
\begin{equation} 
D_{SS}({\bf q},\omega)= \int\, dt\,d{\bf r}\,e^{i\omega t} 
\,e^{-i{\bf q}\cdot{\bf r}} \,
\langle-i\,T\bigg(\sigma({\bf r}, t)\,-\,\langle\sigma\rangle\, ,\,
\sigma(0)\,-\,\langle\sigma\rangle\bigg)\rangle
\end{equation}
We have replaced the retarded Green's function 
by the time-ordered one which is identical for  positive frequencies.  
One can also define a momentum dependent susceptibility according to~: 
\begin{equation} 
\chi_S({\bf q})=2\,{\langle \bar q q\rangle^2_{vac}\over f^2_\pi} \,
Re D_{SS}({\bf q},\omega=0) \label{chiq}
\end{equation}
The range of the fluctuations of the quark density is thus given by the one
of the sigma field. Notice that in the phase of broken symmetry, which is 
the case at ordinary densities, these fluctuations correspond to longitudinal ones, 
{\it i.e.}, along the direction of the spontaneous ordering  which is that 
of the scalar field. In a very simple picture where the sharp sigma mass is 
reduced in the nuclear medium, {\it i.e.} $m_\sigma$
replaced by some dropped value, $m^*_\sigma$, as has been  suggested by
Hatsuda 
{\it et al.} 
\cite{HKS99}, 
the static correlator is $exp(-m^*_\sigma r)/r$.
Hence, as $m^*_\sigma$ goes to zero at full chiral symmetry restoration, the
fluctuations acquire an infinite range as for fluids  
near the critical temperature. 
In the work of Hatsuda {\it et al.} the sigma mass modification arises from the tadpole
term. Here, we include the coupling of the sigma to the nucleon-hole 
excitations, which modifies the scalar field propagator as follows:
\begin{equation}
D_{SS}=D^0_S\,(1\,+\,
D^0_S\,g^2_S\Pi_{SS})\label{DRPA}
\end{equation}
Here  $D^0_{S}=1/(q^2-m^2_{\sigma})$ is the bare sigma propagator,  
$\Pi_{SS}$   is the full scalar $NN^{-1}$ polarization propagator, and  $g_S$
is  the $\sigma NN$ coupling constant. This last quantity  is a
function of the momentum  transfer but it may also be density dependent, as proposed
in the quark meson coupling model (QMC)\cite{G88}. 
We will come back to this question. In the low density limit where
the non-relativistic approximation applies, the polarization propagator $\Pi_{SS}$
reduces to the free Fermi gas expression~:
\begin{equation}
 Re \Pi^0({\bf q},\omega=0)
=-{M_N\,k_F\over \pi^2}\,\left[1+{1\over t}\left(1-{t^2\over 4}\right)\,
ln\left({1+t/2\over 1-t/2}\right)\right]
\end{equation}  
At low momentum, it is a slowly varying function of $t=|{\bf q}|/k_F$~:
\begin{equation}
Re \Pi^0({\bf q},\omega=0)\simeq\ -{2\,M_N\,k_F\over \pi^2}
\left(1-{t^2\over 12}\right)\label{REP}\end{equation} 
This expression, taken at t=0, introduced in the relation \ref{DRPA},   
provides the low density expression for
the unit volume scalar susceptibility  of the infinite nuclear medium:
\begin{equation}
\chi_S =2\,{\langle \bar q q\rangle^2_{vac}\over f^2_\pi} \,
Re D_{SS}({\bf q=0},\omega=0)\simeq\chi_{S,vac}\left(1\,+
{2\,g^2_S\,M_N\,k_F\over
  \pi^2\,m^2_\sigma}\right)
\end{equation}
Notice that, in our definition, the scalar susceptibility is, as the quark 
condensate,  negative, {\it i.e.}, it increases the magnitude of the condensate. 
An in-medium increase of its  absolute value  thus opposes the 
restoration effect. For the actual evaluation  of the RPA correction in the 
lowest order, we take the parameters from Guichon {\it et al.} \cite{GSRT96}: 
$g_S^2/m_{\sigma}^2 =.71\, GeV^{-2}$. This  estimate   leads to
a large modification of the sigma propagator: at normal nuclear matter 
density $\rho_0$~,
$D_S\sim 12\, D^0_S$. Thus we have reasons to believe in a 
sizeable in-medium increase of the scalar susceptibility. This reflects the increase of the
range of the scalar quark density correlator~: the quark density fluctuations, transmitted by
the sigma meson, are relayed by the nucleons, thus increasing their range. Notice that this
effect can be interpreted as arising from an in-medium decrease of the sigma
mass. This mass is a screening one and  not the energy of the pole 
of the sigma propagator at zero momentum. We point out that the mixing with the $NN^{-1}$
excitations does not affect the scalar spectral function at high energy, beyond the two-pion 
threshold, since $\Pi_{SS}$ vanishes in this energy range, which is well above  the nuclear
excitation energies.

\smallskip\noindent
Now, for a more quantitative evaluation, we have to know how  much the full  
$\Pi_{SS}$ 
deviates from the free one~: $\Pi_0$, which 
is a pure nuclear physics problem, on which experimental information can be 
obtained. Especially at normal nuclear matter density the full polarization propagator 
$\Pi_{SS}$ , at zero momentum, is linked to the incompressibility 
factor K of nuclear  matter, the magnitude 
of which is known, even if the exact value is still under investigation: 
\begin{equation}
\Pi_{SS}({\bf q}=0, \omega=0, \rho_0)\,= -\,{9\,\rho_{ 0}\over K}
\end{equation} 
With the currently suggested value, K=230 MeV, which is practically the free Fermi 
gas one~: $K=3 k^2_F/M_N$,  the quantity   $\Pi_{SS}({\bf q}=0, \omega=0)$   
also has the free Fermi gas  value given in eq.(\ref{REP}). 
This supports the previous first order estimate, which led to a large 
 increase of the scalar susceptibility.

\smallskip\noindent
However, to be more precise, we have to discuss also 
in more details the value of the $\sigma NN$ coupling constant entering 
the renormalization factor in eq.(\ref{DRPA}). We mentioned 
that this quantity may depend on the density. This is the case in the 
quark-meson coupling model (QMC) \cite{G88}. This model preserves 
chiral invariance  and its main consequence can be incorporated 
in the linear sigma model. It has been  
introduced to account for the nuclear saturation, which is not accounted for 
in the original  sigma model \cite{CEO02}. In view of its relevance in our 
problem  we enter in some details of this model. 
It  is an extension  of the quantum hadrodynamics model 
\cite{SW86,SW97}, but formulated directly
at the quark level. In QMC  the scalar meson and vector 
mesons couple directly to the quarks inside the 
nucleons, described by a bag model. The crucial ingredient 
of the model is that the nucleon has an internal structure which adjusts 
to the presence of the scalar field. Indeed, under the influence 
of this attractive  field the quark mass is lowered according to~:
$m^*_q=m_q\, -\,g_\sigma^q \langle\sigma\rangle$
where $g_\sigma^q$ is the sigma quark coupling constant.
Accordingly, the valence quark scalar number, which depends on 
the quark mass, also decreases, the quarks 
becoming more relativistic. This effect is directly related to the QCD 
scalar susceptibility of the bag, $\chi_S^{bag}$. Introducing
the scalar charge, $Q_S$, defined as the valence quark scalar number~:
\begin{eqnarray}
Q_S(\langle\sigma\rangle)=\int\,d{\bf r} \left(\langle\bar q q({\bf 
r})\rangle\,-\,
\langle\bar q q\rangle_{vac}\right) 
&=& Q_S(m_q)\,+\,\chi_S^{bag}\, (m_q^*\,-\,m_q)\nonumber\\
&=& Q_S(\langle\sigma\rangle=0)\,-\,g^q_\sigma\,\chi_S^{bag}\,\langle\sigma\rangle
\end{eqnarray}
Now, the scalar charge acting as the source of the 
scalar field, a decrease of the scalar charge  amounts to a  lowering 
of the  $\sigma NN$ coupling constant, when the mean scalar field, 
{\i.e.}, the density,  increases~:
\begin{equation}
g_S(\rho)=g^q_\sigma\, Q_S(\langle\sigma\rangle)
\end{equation}
In QMC this mechanism is responsible for the saturation of nuclear matter.
Alternately the effect can be viewed as the creation of an induced scalar 
charge, $Q_S^{ind}$, due to the presence of the scalar field and proportional, 
for small intensities, to the field,  with~:
\begin{equation}
Q_S^{ind}=-\,g^q_\sigma 
\,\chi_S^{bag}\,\langle\sigma\rangle,\qquad\hbox{per
 nucleon}.
\end{equation}
In the nuclear medium, the presence of the induced charges modifies the 
propagation  equation of the scalar field. Adding the 
intrinsic and the induced sources,
we have  in the static situation the following result in a uniform medium~:
\begin{equation}
\langle\sigma\rangle={g_S\,\rho_S\over
m^2_\sigma\,+\,(g^q_\sigma)^2 \,\chi_S^{bag}\,\rho_S}
\end{equation}
Here $\rho_S$ is the nucleon scalar density and $g_S$ is the value of the 
coupling constant for a vanishing scalar field.
The introduction of $\chi^{bag}_S$, which is positive   amounts to 
an increase of the sigma mass~: 
\begin{equation}
m_\sigma^{*2}(\rho)=m^2_\sigma\,+\,(g^q_\sigma)^2\,
\chi_S^{bag}\,\rho_S
\end{equation}
which counteracts the decrease due to the mixing with $NN^{-1}$ states.
Interpreted as an optical potential for the propagation of the sigma, 
related to the $\sigma N$ amplitude, this internal structure
effect represents a non-Born $\sigma N$ amplitude.
What we had discussed previously was  the influence of the 
Born amplitude (no excitation of the nucleon took place) taken alone on 
the sigma propagation. We  have now to combine the two influences, of the non-Born
and Born, with the result~: 
\begin{equation}
D_S\,=\tilde D^0_S\,\left(1\,+\,g^2_S\,\tilde D^0_S\,\Pi_{SS}\right)
\label{sigprop}
\end{equation}
with $\tilde D^0_S= -  (m_\sigma^{*2}(\rho))^{-1}$, at zero four momentum.
Comparing with the vacuum value $D^0_S$ we have at normal nuclear matter density~: 
\begin{equation}
D_S={m^2_\sigma\over m^{*2}_\sigma}\,D^0_S\,
 \left(1\,+\,{g^2_S\,\rho_ 0 \over m^{*2}_\sigma}\,
{9\over K}\right)
= {m^2_\sigma\over m^{*2}_\sigma}\,D^0_S\,
\left(1\,+\,g_S\, \langle \sigma (\rho_0)\rangle\,{9\over 
K}\right)\label{DS}
\end{equation}
where $\langle \sigma(\rho)\rangle$ is the average sigma field in 
the medium. For a bag radius of $.8$ fm  Guichon et al \cite{GSRT96} give
$g_S \langle\sigma(\rho_0)\rangle\simeq 200$ MeV. The
corresponding increase of the square sigma mass is $ 18\%$. For consistency  
we use also  the value of the model for the incompressibility: K=280 MeV 
(also compatible with the experimental allowed range). With
these values, the enhancement factor of the scalar susceptibility turns 
out to be $\chi_S(\rho_0)/\chi_{S, vac}=6.2$ ,
still a large medium effect. This increase reflects that of the range of 
the quark scalar density fluctuations which, transmitted by the sigma 
field, are relayed by the nucleons.

\section{Pseudoscalar Susceptibility}

This factor applies to the ``parallel'' susceptibility, 
along the order  parameter. In the broken phase,  there  exists a transverse one, 
along the ``perpendicular'' direction. For QCD this is the 
pseudoscalar susceptibility, linked to the fluctuations of the 
pseudoscalar quark density. We define it in such a way that it coincides with the scalar
susceptibility in the restored phase~:
\begin{equation}
\chi_{PS}=2\,\int dt'\,d{\bf r}'\,\Theta(t\,-\,t')\langle\,-i
\left[\bar q\, i\gamma_5\,{\tau_\alpha\over 2}\,q (0) \, ,\, 
\bar q\, i\gamma_5\,{\tau_\alpha\over 2}\,q ({\bf r}' \, \,t')\right]\rangle
\end{equation}
This pseudoscalar susceptibility is related
to the correlator of the divergence of the axial current since~:
\begin{equation}
\partial^\mu {\cal A}^\alpha_\mu (x)= 2 \, m_q\,\bar q \,i\gamma_5\,
{\tau_\alpha\over 2}\,q (x)
\end{equation}
In representations where PCAC holds, which is the case in the linear sigma 
model or in specific representations of the non-linear one, the 
interpolating pion field is
taken to be proportional to the divergence of the axial current according to~:
\begin{equation}
\partial^\mu {\cal A}^\alpha_\mu (x)= f_\pi\, m^2_\pi \Phi^\alpha(x)
\end{equation}
The pseudoscalar susceptibility  $\chi_{PS}$ is then linked to the pion 
propagator, taken at zero momentum and energy~: 
\begin{eqnarray}
\chi_{PS} &=&  {f^2_\pi\, m^4_\pi\over 2\, m^2_q}
\int dt'\,d{\bf r}'\,\Theta(t\,-\,t')\langle\,-i
\left[\Phi^\alpha (0) \, ,\, 
\Phi^\alpha ({\bf r}' \, \,t')\right]\rangle\nonumber\\
&=&  \,{f^2_\pi\, m^4_\pi\over 2\, m^2_q}
Re D_\pi({\bf q}=0 ,\omega=0)\label{PSS}
\end{eqnarray}
Since the factor multiplying the pion propagator can  be written as 
$2\,\langle \bar q q\rangle^2_{vac}/ f^2_\pi$, this equation  is the 
analog of the one for the scalar susceptibility, with the pion replacing 
the sigma. Thus one reaches, in the linear sigma model, a symmetric situation
where the two susceptibilities are governed by the propagators of the two
chiral partners, $\sigma $ and $\pi $. In the medium we denote $S({\bf 
q}\, ,\,\omega)$ (which implicitely depends on the density)
the pion self- energy so that~: 
\begin{equation}
D_\pi({\bf q}=0 ,\omega)= \left[\omega^2\,-\, m^2_\pi\,-\, 
S({\bf q}=0 ,\omega)\right]^{-1}
\end{equation}
The expression of the self-energy  depends on the representation. The one that should
enter here  is the one which applies to the PCAC representation. Its expression 
which has been discussed by Delorme et al \cite{DCE96} is not simple. 
Firstly because the $\pi-N$ scattering amplitude itself has a complicated 
off-shell dependence, with the sign change between the Cheng-Dashen and 
the soft point. In addition, to second order in the nucleon density, it 
contains terms which are specific to this representation and are imposed 
by chiral symmetry. This complexity is however irrelevant for our purpose 
which is to establish a link with the condensate evolution. 
At zero four momentum one has~:
\begin{equation}
Re D_\pi({\bf q}=0 ,\omega=0)=-\,{1\over m^2_\pi\,+\, S(0,0)}
\end{equation}
On the other hand, the evolution with density of the condensate is 
governed by the nuclear sigma commutator: $\Sigma_A/A$ per nucleon, according 
to the exact expression:
\begin{equation}
{\langle \bar q q(\rho)\rangle\over\langle \bar q q\rangle_{vac}}=
1\,-{(\Sigma_A/A)\,\rho\over f^2_\pi\, m^2_\pi}\label{CON}
\end{equation}
The nuclear sigma commutator, which  a priori depends on the density, is 
defined as the expectation value over the nuclear ground state of the commutator 
between the axial charge and its time derivative. 
From  QCD this quantity is also the volume integral of 
the difference between the in-medium condensate and its vacuum value is: 
\begin{equation}
\Sigma_A=2\, m_q\,\int d{\bf r}\,\big(\langle \bar q q 
(\rho)\rangle\,-\,\langle \bar q
q\rangle_{vac}\big),
\end{equation}
hence the relation  \ref{CON}. In the PCAC representation, the nuclear sigma 
commutator also represents the scattering amplitude for soft pions on  
the nuclear medium, $T(0,0)$ (per unit volume), with:
\begin{equation}
{(\Sigma_A/A)\,\rho\over f^2_\pi}=\,T(0, 0)
\end{equation}
This last quantity  is related to the pion self-energy, $S(q,\omega)$ through:
\begin{equation}
T(0, 0)= {S(0,0)\over 1\,+\,S(0,0)/m^2_\pi}
\end{equation}
In this expression the denominator represents the coherent rescattering of 
the soft pion  \cite{E93}.
This distortion factor is needed in order to make the nuclear sigma 
commutator independent on  the representation, as was discussed in ref \cite{DCE96}. 
In fact, in the density dependence of the condensate, the distortion is 
cancelled by many-body terms of the self-energy  but here we do not need 
the explicit form of the   self-energy and it is necessary to include the 
distortion. It is now possible to establish  a  link between the pseudoscalar 
susceptibility and the in-medium condensate by writing the condensate from 
its expression (eq.\ref{CON}) as: 
\begin{equation}
{\langle \bar q q (\rho)\rangle\over\langle \bar q q\rangle_{vac}}=
1\,-\,{T(0, 0)\over m^2_\pi}= {1\over 1\,+\,S(0,0)/m^2_\pi}
\end{equation}
From the expression of the pseudoscalar susceptibility (eq.\ref{PSS}) and using the GOR 
relation, one finally obtains the following relation, which is 
independent of the representation:
\begin{equation}
\chi_{PS}= {\langle \bar q q\rangle_{vac}\over m_q}\,{1\over 
1\,+\,S(0,0)/m^2_\pi}\,
={\langle \bar q q(\rho)\rangle\over m_q}\label{CHIPS}
\end{equation}
The pseudoscalar susceptibility  follows the condensate evolution, {\it i.e.}, 
its magnitude decreases with  density, with a linear dependence in the 
dilute limit where the relation $(\Sigma_A/A)=\Sigma_N$
holds. At normal density the susceptibility  has thus decreased by 
$35\%$. The presence of the quark mass in the denominator  
of the expression of $\chi_{PS}$ makes it
divergent in the chiral limit, as it should.

\smallskip\noindent
Our previous relation (\ref{CHIPS})  between the transverse (pseudoscalar) 
susceptibility and the order parameter (the condensate) can be understood 
from the magnetic analogy. The rotational symmetry is
intrinsically broken by a magnetic field ${\bf  H}_0$ which aligns the 
spontaneous magnetization along its direction. The application of a small transverse
field ${\bf H}_\perp$  rotates the magnetization  ${\bf M}$ by a small 
angle $\theta$, such that it is now aligned in the direction of the resulting
field ${\bf H}_0 + {\bf H}_\perp$. The transverse magnetization is 
$ M_\perp= M~\theta=  M ( H_\perp/ H_0)$ 
and the transverse susceptibility is  $\chi_\perp=  M_\perp / H_\perp=
M/ H_0$, which is the analog of our formula (\ref{CHIPS}).

\smallskip\noindent              
Since the pseudoscalar susceptibility is governed by the 
pion propagator it is natural to discuss also its relation to the 
pion mass evolution  \cite{CEW96}. 
We define, as usual, the in-medium effective mass, $m^*_\pi(T,\rho)$,
as the energy of the pole of the pion propagator taken at zero 
three-momentum, which differs from the definition of Rajagopal
 and Wilzcek\cite{RW93}, where $m^{*2}_\pi(T,\rho)$ is
  directly taken as the inverse susceptibility.
The problem of 
the relation between the pion mass and the condensate evolutions  has 
already been addressed, for the nuclear medium or the heat bath 
\cite{CEW96}.
In the heat bath for instance the presence of the residue in the pion 
propagator makes the 
inverse squared pion mass to decrease three times more slowly with 
temperature than the condensate. 
A similar difference between the two evolutions occurs in the nuclear 
medium. The density dependence of the quark condensate, the pion 
propagator and the pion mass were extensively studied by Delorme et al
\cite{DCE96}, up to second order in the density, in two  different 
representations of the non-linear sigma model. They have shown 
that both the 
condensate and the pion mass are independent of the representation, as it 
should, the pion propagator instead is not.  We reproduce in table I their 
result for these three quantities in the representation  where PCAC 
applies and in the Weinberg representation. In the first one, the
pion propagator  at zero momentum and  the condensate are proportional. 
The table clearly displays the difference between the inverse pion mass squared and the
pseudoscalar susceptibility evolutions. 

\section{results and conclusion}

We  now comment our conclusion that at normal 
density the scalar and pseudoscalar susceptibilities become closer than 
in the vacuum. In order to show that this convergence is not accidental 
but is a manifestation of chiral symmetry restoration, we have to enter 
into  
the various mechanisms responsible for the restoration process.
One of them is the nuclear scalar field melting into the condensate, 
which has the same quantum number (see fig.1). It leads to a modification of the 
condensate, $\Delta_\sigma \langle \bar q q(\rho)\rangle$ (accordingly of the 
pseudoscalar susceptibility) by : 
\begin{equation}
\Delta_\sigma \langle \bar q q(\rho\rangle
= - \langle \bar q q(vac)\rangle \,\frac{g_s\rho_s}{f_{\pi}m_{\sigma}^2}
=B\rho_s,
\end{equation}
in which the second equation 
simply defines the proportionality factor B.
Now, comparing this to the modification of the scalar suceptibility
arising from the coupling of the scalar field to the nucleons, which can be written
as $\Delta \chi_S=2\,B^2\,\Pi_{SS}$,
we see that the factor B enters with a square 
power 
and the nucleon 
density has been replaced by the nucleon-hole propagator. But a unique  
mechanism has
produced the decrease of the order parameter and the increase of the 
scalar susceptibility. A similar correspondence exists
 for the influence of 
the nuclear pions. Their scalar density, which is linked to the mean value 
of the square pion field, $<\Phi^2>$,
modifies the condensate according to:
\begin{equation}
\Delta_\pi \langle \bar q q(\rho \rangle
= - \langle \bar q q(vac)\rangle \,\frac{\langle\Phi^2 \rangle}{2f_{\pi }^2}
=C\,\langle\Phi^2 \rangle\label{PHI}
\end{equation}
While the influence of the two-pion continuum on the scalar susceptibility is:
\begin{eqnarray}
\Delta\chi_S & =& 2\,{\langle \bar q q(vac)\rangle^2\over f^2_\pi}\,{1\over
m^4_\sigma}\,{m^4_\sigma\over 4 f^2_\pi}\,G_{2\pi}({\bf q}=0 , \omega=0)\nonumber\\
& =& 2\,C^2\, G_{2\pi}({\bf q}=0 , \omega=0)\label{LINK}
\end{eqnarray}
In the scalar susceptibility  the 
pion density is replaced by the two-pion propagator 
$G_{2\pi}$ and the factor C enters at the square power. The 
evolutions of the two  susceptibilities are correlated in the sense that 
a unique mechanism (for instance the melting of the scalar 
nuclear field in the condensate) is responsible for  the enhancement of 
one susceptibility and the decrease of 
the other. In fact it is possible to establish this link by a direct  
evaluation of the scalar susceptibility from the condensate, by taking the 
derivative with respect to the quark mass. Consider for instance  the 
contribution from the pion loops, which modifies the condensate according 
to  equation \ref{PHI}. The nuclear pion density, $\langle\Phi^2 \rangle $, 
depends on the pion mass, {\it i.e.}, 
on the quark mass and thus generates a contribution to the 
nuclear scalar susceptibility. Since $\langle\Phi^2 \rangle $ is related 
to the pion propagator, its derivative with respect to $m_q$ 
is linked to the two-pion 
propagator taken at zero momentum, which leads to our previous equation \ref{LINK}.

\smallskip\noindent
Coming back to the convergence of the two
susceptibilities, it arises from both evolutions. The smaller relative  decrease of 
the pseudoscalar one is compensated by the large value of this susceptibility, 
owing to the smallness of the pion mass. In view of this large  convergence effect
at normal density, it is natural to explore the phenomenon at larger densities. 
We cannot  perform this extrapolation with certainty but we can have some 
indications. The mixing of the $NN^{-1}$ states with the sigma which is 
responsible for the increase of the scalar susceptibility may not develop 
much further for several reasons. The quantity $\Pi_{SS}$  involves the scalar 
nucleon density which  increases more slowly than  the ordinary density.  
Moreover it is proportional to the effective nucleon mass which decreases 
with density.  Finally the nucleon reaction to the scalar field manifests 
itself more with  increasing density. In order to evaluate the influence of 
these effects at any density, we  take for the quantity $g_S^2\,\tilde 
D_0^S\,\Pi_{SS}({\bf 
q}=0, \omega=0)$ the ansatz:
\begin{equation}
g_S^2\,\tilde 
D_0^S\,\Pi_{SS}=g_S\,\langle\sigma(\rho)\rangle\,{3\,M^*_N(\rho)\over k^2_F}
\end{equation} 
which holds at $\rho_0$, since K is close to the Fermi gas value, but may 
not when the density increases. 
We take  the values of the scalar field and of the effective nucleon mass from 
\cite{GSRT96}. With these inputs we find that the enhancement factor of 
the scalar susceptibility stabilizes, with even a certain decrease. 
It is 5.2  at  $1.5\rho_0$ and 4.8    at $2\rho_0$. 
The behavior with density of the two susceptibilities is  shown in fig.2
but we stress again that for the scalar one, only the point at $\rho_0$
rests on the experimental input of the compressibility. Moreover this 
evaluation only takes into account the mixing of the sigma with the 
nucleon-hole states. Its mixing with two-pion states should 
also be incorporated especially at large densities.
The point at $.5\rho_0$, evaluated with the same ansatz, is only given for 
illustration as one enters here in the region of the spinodal instability.

\bigskip\noindent 
In summary we have studied  two QCD susceptibilities of the nuclear medium, 
the scalar-isoscalar  one and the pseudoscalar-isovector one. They are
linked to the fluctuations of the corresponding quark densities. For the 
first one, the use of the linear sigma model provides a link 
with the propagator of the sigma meson. In the nucleus this meson mixes 
with the low lying scalar-isoscalar nuclear excitations. In this respect, 
it is important to stress
the following point. We have assimilated the sigma field of the linear sigma model 
({\it i. e.} the chiral partner of the pion) with the  nuclear scalar field
responsible of nuclear binding in QHD. However the latter is a chiral 
invariant while
the chiral partner of the pion is not. We have shown in a previous work \cite{CEG02} 
that the two fields can be
related in the common framework of the linear sigma model. They differ
by a term proportional to the pion scalar density, which has a quenching  effect  on
the fluctuations. This is presently ignored.  It
will be taken into account in a forthcoming work, 
but we expect a moderate influence.   One can also question the
relevance of the sigma model for the problem of the quark density fluctuations. 
Actually these
fluctuations  couple to those of the nucleon scalar
density, which increases their range. The linear sigma model
 provides an evaluation for this coupling which must be present. At the normal 
nuclear density the mixing with the nuclear excitations is constrained by 
standard nuclear phenomenology,
{\it i.e.}, by the nuclear matter incompressibility. It leads to 
 an increase of the magnitude of the scalar 
susceptibility  by a factor of about 6. This effect, although 
pronounced at normal density, does not appear to increase further  with
increasing density. Indeed, at higher densities, the sigma is expected 
to decouple from 
the  scalar $NN^{-1}$ excitations. As for the  pseudoscalar susceptibility,
which is linked to the pion propagator, we have shown that 
it follows the evolution of the condensate, {\it i.e.}, its magnitude 
decreases with density. The two combined effects make 
the scalar and pseudoscalar susceptibilities appreciably closer in the 
nuclear medium  than in  the vacuum, already  at the normal  density. 
This convergence is a  strong  signal of chiral symmetry restoration.

\bigskip
Acknowledgments~: We have benefited from fruitful discussions 
with W.Alberico, D. Davesne, P. Guichon, A. Molinari and J. Wambach. 

\newpage
\begin{table}
\begin{center}
\caption{Comparison of the results obtained with two non-linear Lagrangians,
 ${\cal L}^W$ which does not satisfies PCAC and ${\cal L}'$ which does.
The successive lines give the inverse pion
propagator, the squared pion effective mass and the density
evolution of the quark condensate, {\it i.e.}, of the pseudoscalar susceptibility.
We define $x =\rho\Sigma_N/f_\pi^2 m_\pi^2$  and
$x_{2,3} = \rho\, c_{2,3}/f_\pi^2\;$ where $c_{2,3}$ are the standard 
parameters of the chiral Lagrangian [12]. }      

\label{tab1}
\medskip
\begin{tabular}
 {|c||c|c|}  \hline
& & \\
 & ${\cal L}^W$ &$ {\cal L}' $   \\ \hline \hline 
 &  &  \\ 
$ D_\pi ^{-1} (\omega, {\hbox{\boldmath $q$\unboldmath}} \,) $ &
 $\big [ 1+2(x_2+x_3) \big]\omega^2$
& $ \bigg[\big [1+2(x_2+x_3)\big]\omega^2
 - (1+2x_3){\hbox{\boldmath $q$\unboldmath}}^2$ \\
 &\hfill $ -  (1+2x_3){\hbox{\boldmath $q$\unboldmath}}
^2 - (1-x)m_\pi^2$ &\hfill $  -
 (1-x)m_\pi^2 \bigg] /(1-x)^2 $ \\ 
& & \\
\hline 
 &  &  \\ 
$\displaystyle{\frac{ m_\pi^{\ast 2}}{m_\pi^2}}$ &
 $\displaystyle{\frac{1-x}{1+2(x_2+x_3)}}$ & 
$ \displaystyle{\frac{1-x}{1+2(x_2+x_3)}} $\\
& & \\
\hline 
 &  &  \\ 
$\displaystyle{\frac{< \overline {q}q (\rho) >} { < \overline {q}q (0) >}}$
 & $1-x$ & $1-x $ \\
& &  \\ \hline
\end{tabular}
\end{center}
\end{table}
\vglue 1 true cm
\begin{figure}[H]
\begin{center}
\epsfig{file=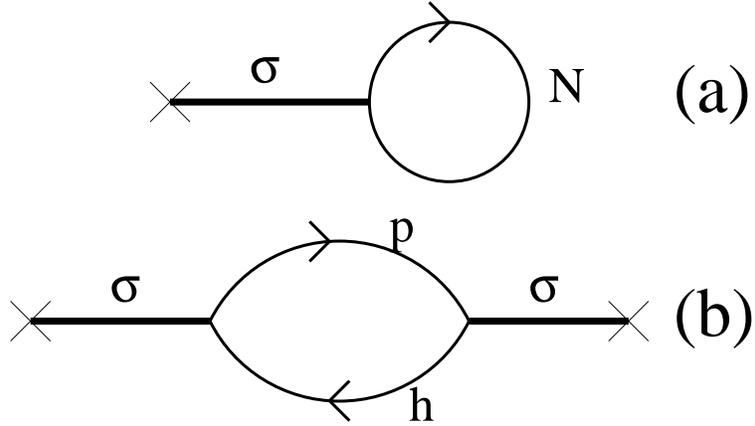,width=10.0cm,angle=0}
\end{center}
  \caption{ Influence of the melting  of the scalar field into the condensate  
  (represented by a cross) on (a) the condensate, (b) the scalar susceptibility.}
\label{CHI}      
\end{figure} 
\vspace{1. cm} 
\begin{figure}[H]
\begin{center}
\epsfig{file=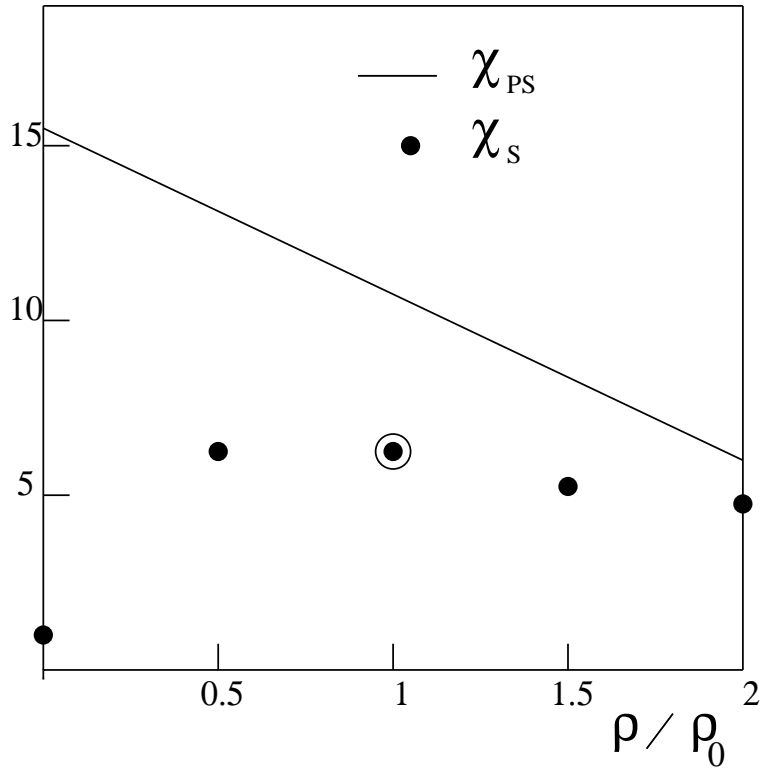,width=10.0cm,angle=0}
\end{center}
  \caption{Pseudoscalar (continuous line) and scalar (dots)
susceptibilities as a function of the nuclear density. Both are normalized to the
vacuum scalar susceptibility with a sigma mass of 550 MeV. In the pseudoscalar
susceptibility the independent nucleon approximation has been assumed for the quark
condensate evolution. For the scalar one, the points has been evaluated with the ansatz
described in the text, but for the normal density point in which the 
``experimental'' incompressibility has been introduced (its special character is
indicated by the double circle).}
\label{CHI}      
\end{figure}  
\newpage



\begin{thebibliography}{99}
\bibitem{CSSS95} L.S. Celenza, C.M. Shakin, W.D. Sun and J. Szweda, Phys. 
Rev C51 (1995)937
\bibitem{DZCW00} D. Davesne, Y.J. Zhang, G. Chanfray and J. Wambach
nucl-th/0002032, Proceedings of the Hirschegg 2000 workshop, (2000) 88.
\bibitem{ARGCW95} Z. Aouissat, R. Rapp, G. Chanfray, P. Schuck and J.
Wambach, Nucl. Phys.. A581 (1995) 471.
\bibitem{HKS99} T. Hatsuda, T. Kunihiro and H. Shimizu, Phys. Rev. Lett.82 (1999) 2840.
\bibitem{WAS01} J. Wambach, Z. Aouissat and P. Schuck, Nucl. Phys. A690 (2001) 127c.
\bibitem{HK85} T. Hatsuda and T. Kunihiro, Phys. Rev. Lett. 55 (1985) 158.
\bibitem{RW93} K. Rajagopal and F. Wilczek, Nucl. Phys. B399 (1993) 395.
\bibitem{G88} P.A.M.Guichon Phys. Lett.B200 (1988) 235.
\bibitem{GSRT96} P.A.M. Guichon, K.Saito, E.Rodionov and  A.W. Thomas, 
Nucl.Phys. A601 (1996) 349.
\bibitem{CEO02} G. Chanfray, M. Ericson and M. Oertel, work in progress.
\bibitem{SW86} B.D. Serot and J.D. Walecka, Adv. Nucl. Phys. 16 (1986) 1.
\bibitem{SW97} B.D. Serot and J.D. Walecka, Int. J. Mod. Phys. E16 (1997) 515.
\bibitem{DCE96}J. Delorme, G. Chanfray and M. Ericson, Nucl. Phys. A603 
(1996) 239.
\bibitem{E93} M. Ericson, Phys Lett. B101 (1993) 11.
\bibitem{CEW96} G. Chanfray, M. Ericson and J. Wambach, Phys. Lett. B388 (1996) 673.
\bibitem{CEG02} G. Chanfray, M. Ericson and P.A.M.Guichon, Phys. Rev C63 055202.
\end{thebibliography}
\end{document}